
\input harvmac

\def\a{\rightarrow}

\def\cc{\cosh^2\alpha}
\def\c2{\cosh 2\alpha}
\def\d{\nabla}
\def\l{\lambda}
\def\m{M_0}
\def\r{\rho}
\def\ss{\sinh^2\alpha}
\def\s2{\sinh 2\alpha}
\def\t{\tilde}
\def\RN{Reissner-Nordstr{\o}m}
\def\({\left(}
\def\){\right)}
\def\frac#1#2{{#1 \over #2}}
\def\[{\left[}
\def\]{\right]}
\gdef\journal#1, #2, #3, 19#4#5{
{\sl #1~}{\bf #2}, #3 (19#4#5)}

\lref\banks{T. Banks, A. Dabholkar, M. Douglas, and M. O'Loughlin,
``Are Horned Particles the Climax of Hawking Evaporation?",
\journal Phys. Rev., D45, 3607, 1992; T. Banks and M. O'Loughlin,
``Classical and Quantum Production of Cornucopions at Energies Below
$10^{18}$ GeV", Rutgers preprint RU-92-14, hep-th/9206055;
T, Banks, M. O'Loughlin, and A. Strominger, ``Black Hole Remnants and the
Information Puzzle", Rutgers preprint RU-92-40, hep-th/9211030.}
\lref\bekenstein{J. Chase, \journal Commun. Math. Phys., 19, 276, 1970;
J. Bekenstein, \journal Phys. Rev., D5, 1239, 1972.}
\lref\brill{D. Brill, ``Splitting of an Extremal \RN\ Throat via Quantum
Tunneling", \journal Phys. Rev., D46, 1560, 1992.}
\lref\brpf{D. Brill and H. Pfister, ``States of Negative Total Energy in
Kaluza-Klein Theory", \journal Phys. Lett., B228, 359, 1989;
D. Brill and G. Horowitz, ``Negative Energy in String Theory",
\journal Phys. Lett., B262, 437, 1991.}
\lref\ghs{ D.~Garfinkle, G.~Horowitz and A.~Strominger,
``Charged Black Holes in String Theory,''
\journal Phys.~Rev., D43, 3140, 1991; {\bf D45}, 3888({\bf E}) (1992).}
\lref\garfinkle{D. Garfinkle, private communication.}
\lref\gima{ G. Gibbons, ``Antigravitating Black Hole
Solitons with Scalar Hair in N=4 Supergravity", \journal Nucl. Phys., B207,
337, 1982;
G.~Gibbons and K.~Maeda, ``Black Holes and Membranes in
Higher-Dimensional Theories with Dilaton Fields",
\journal Nucl.~Phys., B298, 741, 1988.}
\lref\grha{R. Gregory and J. Harvey, ``Black Holes with a Massive Dilaton",
Enrico Fermi Preprint EFI-92-49, hep-th /9209070.}

\lref\harrison{B. Harrison, ``New Solutions of the Einstein-Maxwell Equations
from Old", \journal J. Math. Phys., 9, 1744, 1968.}
\lref\hast{J. Harvey and A. Strominger, ``Quantum Aspects of Black Holes",
to appear in the proceedings of the 1992 Trieste Spring School on String
Theory and Quantum Gravity, hep-th/9209055.}
\lref\hawking{S.~Hawking,
\journal Phys. Rev. Lett., 26, 1344, 1971.}
\lref\howi{ C.~Holzhey and F.~Wilczek,
``Black Holes as Elementary Particles,'' \journal Nucl. Phys., B380, 447,
1992.}
\lref\hohomass{J.~Horne and G.~Horowitz, ``Black Holes Coupled to a Massive
Dilaton", Nucl. Phys. B to appear, hep-th/9210012.}
\lref\horowitz{G. Horowitz, ``The Dark Side of String Theory: Black Holes and
Black Strings",
to appear in the proceedings of the 1992 Trieste Spring School on String
Theory and Quantum Gravity, hep-th/9210119.}
\lref\israel{W. Israel, ``Event Horizons in Static Electrovac Spacetimes",
\journal Commun. Math. Phys., 8, 245, 1968.}
\lref\klopp{R. Kallosh, A. Linde, T. Ortin, A. Peet, and A. Van Proeyen,
``Supersymmetry as a Cosmic Censor", Stanford preprint SU-ITP-92-13;
R. Kallosh, T. Ortin, and A. Peet, ``Entropy and Action of Dilaton Black
Holes", Stanford preprint SU-ITP-92-29, hep-th/9211015.}
\lref\mapa{S. Majumdar, \journal Phys. Rev., 72, 930, 1947; A. Papepetrou,
\journal Proc. Roy. Irish Acad., A51, 191, 1947; J. Hartle and S. Hawking,
\journal Commun. Math. Phys., 26, 87, 1972.}
\lref\psstw{ J.~Preskill, P.~Schwarz, A.~Shapere, S.~Trivedi,
and F.~Wilczek,
``Limitations on the Statistical Description of Black Holes,''
\journal Mod. Phys. Lett., A6, 2353, 1991.}
\lref\sen{S.~Hassan and A.~Sen,
``Twisting Classical Solutions in Heterotic String Theory,''
\journal Nucl. Phys., B375, 103, 1992;
J. Maharana and J. Schwarz, ``Noncompact Symmetries in String
Theory, Caltech preprint CALT-68-1790, hep-th/9207016.}
\lref\senrev{A. Sen, `` Black Holes and Solitons in String Theory",
Tata preprint TIFR-TH-92-57, hep-th/9210050.}

\Title{\vbox{\baselineskip12pt\hbox{UCSBTH-92-52}
\hbox{hep-th/9301008}}}
{\vbox{\centerline {What is the True Description of}
       \centerline {Charged Black Holes?}}}
\centerline{{ Gary T. Horowitz}\footnote{$^*$}
{To appear in the Brill Festschrift, B. L. Hu and T. A. Jacobson, eds.,
Cambridge University Press, 1993.}}
\vskip.1in
\centerline{\sl Department of Physics}
\centerline{\sl University of California}
\centerline{\sl Santa Barbara, CA 93106-9530}
\centerline{\sl gary@cosmic.physics.ucsb.edu}
\bigskip
\centerline{\bf Abstract}
If string theory describes nature, then charged black holes are not described
by the \RN\ solution. This solution must be modified to include
a massive dilaton. In the limit of vanishing dilaton mass,
the new solution can be found by a generalization of the Harrison
transformation
for the Einstein-Maxwell equations. These two  solution generating
transformations and the resulting black holes are compared. It is shown
that the extremal black hole with massless
dilaton can be viewed as the ``square root" of the extremal \RN\
solution. When the dilaton mass is included, extremal black holes are
repulsive,
and it is energetically favorable for them to bifurcate into smaller holes.

\Date{12/92}

\newsec{Introduction}

It is a pleasure to honor Dieter Brill on the occasion of his sixtieth
birthday. Over the years, Dieter has worked on many aspects of
general relativity. But two of his recent interests are negative
energy (in higher dimensional theories) \brpf, and the possibility that
extremal
charged black holes can quantum mechanically bifurcate \brill. I would like to
describe some recent work which touches on both of these areas.

For many years, it has been widely believed
that static charged black holes in nature are
accurately described by the \RN\ solution. This is the result of two
powerful theorems: the uniqueness theorem \israel\
which proves that the only static
black hole solution to the Einstein-Maxwell equations is \RN, and the
no-hair theorem \bekenstein\
which shows that if one adds simple additional matter fields
to the theory,
the only static black hole is still \RN. The extra fields either fall into the
hole or radiate out to infinity. Although this solution seems physically
reasonable in most respects, its extremal limit
has two puzzling features. The first is that the event horizon is infinitely
far away along a spacelike geodesic, but only a finite distance away along
a timelike or null geodesic. The second is that
the extremal black hole has zero  Hawking temperature, but  nonzero entropy
as measured by its surface area.

Recently, a slightly different picture of charged black holes has emerged.
This is the result of studying gravitational consequences of string theory.
A new class of charged black hole solutions have been found\foot{For
recent reviews, see \horowitz \senrev.} which do
not have the puzzling features of \RN.
In the extremal limit, the area of the event
horizon of these new black holes goes to zero, which is consistent
with zero entropy. From another standpoint (for a magnetically charged
black hole) the horizon moves off to infinity in all directions:
timelike, spacelike, and  null. These solutions should be of particular
interest to Dieter Brill since we will see that they  are even more likely to
bifurcate
than the analogous black holes in the Einstein-Maxwell theory. In addition,
although the
total energy remains positive for these four dimensional holes, we
will see that some of them resemble negative energy objects in that
they are repulsive.

In addition to the metric $g_{\mu\nu}$ and Maxwell field $F_{\mu\nu}$,
string theory predicts the existence of
a scalar field called the dilaton. The dilaton has an exponential coupling
to $F^2$ which implies that it cannot remain zero when $F^2 \ne 0 $.
This is the key difference from the previous no-hair theorems  and shows
that the charged black holes will differ from the
\RN\ solution.  Although
the dilaton is massless while supersymmetry is unbroken,  it is expected
to become massive at low energies. The breaking of
supersymmetry in string theory
is a nonperturbative effect which is not well understood at
this time.
So we cannot yet calculate reliably the dilaton potential from first
principles.
However the qualitative behavior of the solutions can be obtained by
considering the simplest possibility $m^2 \phi^2$. We are thus led to
consider the low energy action:
\eqn\lowenergy{  S=\int d^4x \sqrt{-g}
   \left[ R - 2(\nabla\phi)^2  -2m^2\phi^2
		 - e^{-2\phi} F^2 \right]}
with equations of motion
\eqna\eom
$$\eqalignno{    &\d_\mu (e^{-2\phi}F^{\mu\nu}) =0 \>, & \eom a \cr
      &\d^2 {\phi}- m^2 \phi =- \half e^{-2\phi}F^2 \>,& \eom b \cr
    & G_{\mu\nu}  = 2 \d_\mu \phi  \d_\nu \phi +
	   2 e^{-2\phi} F_{\mu\rho} {F_\nu}^\rho \cr
       &- g_{\mu\nu} [(\nabla\phi)^2 + m^2 \phi^2 + \half e^{-2\phi} F^2]
							    & \eom c \cr} $$
When $F_{\mu\nu} =0$, these equations reduce to those of the Einstein-massive
scalar field theory. The no-hair theorems show that the only black
hole solutions are Schwarzschild with $\phi =0$. However, as we just
remarked, when the
charge is nonzero,  the dilaton will not be constant, and will alter the
geometry.
It does not appear possible to write the exact solution to \eom{}\
describing a static,
spherically symmetric, charged black hole  in closed form. But
we will see that by combining general arguments, approximation methods, and
numerical results, we can obtain
a  fairly complete understanding of their properties.

\newsec{Solutions with a Massless Dilaton}

Let us begin by considering the limit where the dilaton mass is negligible.
It will be shown
in the next section that this is appropriate for small black holes.
Even with $m=0$, the exponential coupling of the dilaton
to the Maxwell field appears to prevent a simple closed form solution. However
exact solutions can be found \gima\ghs\
which are, in some respects, even simpler than
the \RN\ solution. In the Einstein-Maxwell theory there is a transformation,
found by Harrison \harrison,
which maps stationary vacuum solutions into stationary
charged solutions. It turns out that there is a similar
transformation for the theory \lowenergy\ with $m=0$ \sen.
In this section, we compare these two transformations and the resulting
charged black holes.
For simplicity, we will consider only static solutions.

We start with the familiar Einstein-Maxwell theory. Suppose
\eqn\start{ ds^2 = -\l dt^2 + h_{ij} dx^i dx^j}
is a static vacuum solution. The Harrison transformation begins by
rewriting this metric as
\eqn\rewt{  ds^2 = -\l dt^2 + \l^{-1} \gamma_{ij} dx^i dx^j}
where $\gamma_{ij} = \l h_{ij}$. The charged solution is obtained by
keeping $\gamma_{ij}$ unchanged and setting
\eqna\newem $$\eqalignno{\t\l =& {\l \over (\cc -\l \ss )^2 } &\newem a \cr
 A_t =&  {(\l-1)\s2 \over 2(\cc - \l\ss )}  &\newem b }$$
where $\alpha$ is a free parameter. Clearly, $\alpha = 0 $ leaves the solution
unchanged.
Since the factor in the denominator appears frequently, it is convenient
to introduce the notation
\eqn\defphi{e^{-2\phi} \equiv \cc -\l \ss}
Notice that $\phi$, at this point,  has nothing to do with the dilaton.
We are considering solutions to the Einstein-Maxwell equations and
$\phi$ is simply defined by \defphi. Equation \newem{a}\ can now be rewritten
$\t\l = \l e^{4\phi}$ and
the new metric can be expressed
\eqn\newsoln{ds^2 = -\l e^{4\phi} dt^2 + e^{-4\phi} h_{ij} dx^idx^j }

Let us apply this transformation to the Schwarzschild solution:
\eqn\schwar{\l = 1 - {2M_0 \over \r}, \qquad h_{ij} dx^i dx^j =
  \(1- {2M_0 \over \r} \)^{-1} d\r^2
  + \r^2 d \Omega,}
In this case, $e^{-2\phi} = 1 + 2\m\ss/\r$. So the new metric becomes
\eqnn\rnhope $$\eqalignno{ ds^2 =&-\(1-{2M_0 \over \r}\)
 \(1 + {2\m\ss \over \r}\)^{-2} dt^2 +  \cr
  &\(1 + {2\m\ss \over \r}\)^2 \[\(1 - {2M_0 \over \r}\)^{-1} d\r^2
+ \r^2 d\Omega\] &\rnhope}$$
This can be simplified by introducing a new radial coordinate $r = \r +2\m\ss$.
Then
\eqn\relate{\r(\r-2M_0) =
(r-2\m\ss)(r-2\m\cc) \equiv (r-r_-)(r-r_+)}
where we have defined
$ r_+ \equiv 2\m\cc$ and $r_- \equiv 2\m \ss$. The metric \rnhope\
now becomes:
\eqn\rnyes{ ds^2 = -{(r - r_+)(r-r_-) \over r^2} dt^2 + {r^2 \over
 (r - r_+)(r-r_-)} dr^2 + r^2 d\Omega }
This is the familiar form of the \RN\ solution with the event horizon
at $r=r_+ $ and the inner horizon at
$r= r_-$. Notice that the original Schwarzschild singularity $\r = 0$
has become the
inner horizon, while the \RN\ singularity $r=0$  corresponds to a negative
value of $\r$ in the original Schwarzschild metric.
The Maxwell field is
\eqn\maxwell{ F_{rt} = {\m \s2 \over r^2}}
which shows that the total charge is $Q = \m\s2$.
The total mass is $M= (r_+ + r_-)/2 = \m\c2$. The extremal
limit is obtained by taking $\alpha \a \infty, \m\a 0$ keeping $\m e^{2\alpha}$
finite. In this limit, the two horizons coincide $r_+ = r_-$.
The event horizon becomes
degenerate, with nonzero area but vanishing Hawking temperature.

Now we turn to the theory with a massless dilaton. There is a solution
generating transformation directly analogous to the Harrison transformation.
It is usually discussed in terms of the conformally rescaled metric
$\hat g_{\mu\nu} = e^{2\phi} g_{\mu\nu}$ (where $\phi$ is now the dilaton)
which is called the string metric.
This is the metric that
a string directly couples to. To avoid confusion, the metric we have been
using until now will be called the Einstein metric (since it has the standard
Einstein action). If one  starts with a static vacuum solution
with (string) metric of the
form $\widehat{ds}^2 = -\l dt^2 + h_{ij}dx^i dx^j$ and $\phi=0$,
then a new solution
is obtained
by keeping $h_{ij}$ unchanged, and setting
\eqna\newemag $$\eqalignno{ \t\l =& {\l \over (\cc -\l \ss)^2} &\newemag a \cr
 A_t =& {(\l-1)\s2 \over 2\sqrt 2(\cc - \l\ss)} &\newemag b \cr
e^{-2\phi} =& \cc - \l\ss &\newemag c} $$
The similarity to \newem{}\ is remarkable.
The change in $\l$ is identical to the Harrison
transformation, and the expression for the vector potential
$A_t$ differs only
by a factor of $\sqrt 2$. In addition, the solution for the dilaton
is exactly what
we called $\phi$ in \defphi. (This is why we chose that definition
for $\phi$.) To compare with the \RN\ solution, we should rescale back
to the Einstein metric:
\eqn\newstr{ds^2 =  -\t\l e^{-2\phi} dt^2 + e^{-2\phi} h_{ij} dx^i dx^j}
In terms of the original $\l$ we get
\eqn\newfin{ ds^2 =  -\l e^{2\phi} dt^2 + e^{-2\phi} h_{ij} dx^i dx^j}
If one now compares  \newfin\
with the solution to the Einstein-Maxwell theory \newsoln,
one sees that {\it the only difference in the solutions with and without a
massless dilaton is
a factor of two in the exponent.}

Applying this transformation to the Schwarzschild solution \schwar, we obtain
the metric
\eqnn\strbh  $$\eqalignno{ds^2 =& -\(1-{2\m\over \r}\)
   \(1+{2\m\ss\over \r}\)^{-1} dt^2 +   \cr
&\(1+{2\m\ss\over \r}\)\[\(1-{2\m\over \r}\)^{-1} d\r^2 +\r^2 d\Omega
\]  &\strbh } $$
together with the Maxwell field and dilaton
\eqna\strmax $$\eqalignno{
A_t =&- {\m\s2 \over \sqrt 2(\r + 2\m \ss)} &\strmax a  \cr
  e^{-2\phi} =& 1+ {2\m\ss \over \r} &\strmax b \cr} $$
The factor of two difference in the exponent has the following
important  consequence:
The surface $\r = 0 $ in the metric \strbh\ still has zero area
and does not become an inner horizon.
The total mass is now $M=\m\cc$ and the charge is $Q=\m\s2 /\sqrt 2$.
Introducing the same radial coordinate as before, $ r = \r + 2\m\ss$,
the solution takes the remarkably simple form
$$
ds^2 = - \(1-\frac{2M}{r}\) dt^2 + \(1-\frac{2M}{r}\)^{-1} dr^2
+ r\(r-\frac{Q^2}
{M} \)d\Omega $$
\eqn\einmet{ F_{rt}=\frac{Q}{r^2}\qquad e^{2\phi} = 1-\frac{Q^2}{Mr}}
Note that the metric in the $r-t$
plane is identical to Schwarzschild! (Although the mass is different from the
one we started with.) There is an event horizon at $r=2M$
and no inner horizon. The only difference from Schwarzschild is that the
area of the spheres is reduced by an amount depending on the charge. This
area goes to zero when $r=Q^2/M$ ($\r=0$) resulting in a curvature singularity.

Clearly,
the solution \einmet\ describes a black hole only when the singularity is
inside
the event horizon i.e. $Q^2 < 2M^2$. Linearized perturbations about this
solution have been studied \howi\ and remain well behaved outside the
horizon: the black hole
is stable. In the extremal limit, $Q^2 = 2M^2$,
the event horizon shrinks down to zero area and becomes singular. The
resulting spacetime describes
neither a black hole (with a spacelike singularity),
nor a conventional naked singularity (which is timelike). Since
the causal structure in the $r-t$ plane is independent of $Q$, the extremal
black hole has a null singularity. Its Penrose diagram is identical to the
region $r>2M$ of Schwarzschild, with both the future and past horizons,
$r=2M$, replaced by singularities.

The fact that the area of the event horizon
goes to zero in the extremal limit has two important consequences.
First, it follows immediately from the area theorem that there is no
classical process by which a nearly extremal black hole can become extremal
\garfinkle.
This is also true for the \RN\ solution, but the proof is more involved.
Second, if one interprets the area  as a measure of the entropy of a black
hole through the Hawking-Bekenstein formula, $S=A/4$, then the extremal black
holes have zero entropy.  This is what one might expect for a ground state.
The nonzero area of the extremal \RN\ solution  has always been puzzling
from this standpoint, since it
seems to indicate a highly degenerate ground state. Although this black
hole resolves one puzzle, it creates a new one.
For a static, spherically symmetric black hole, the Hawking temperature
can be found by analytically continuing to imaginary time and computing
what periodicity is necessary to avoid a conical singularity. Since this
only involves the $r-t$ part of the metric which is identical to Schwarzschild
for the solution \einmet,
the Hawking temperature is also identical: $T= 1/8\pi M$,
independent of $Q$! In particular, the Hawking temperature does not go
to zero as one approaches the extremal limit. This leads one to worry that
the black hole might continue to evaporate past the extremal limit to
form a true naked singularity.
However, the backreaction clearly becomes important as one approaches
the extremal limit, and a complete calculation is necessary  to understand
the evolution\foot{For a review of recent work in this area, see \hast.}.

In terms of the original Schwarzschild mass $\m$ and transformation parameter
$\alpha$, the extremal limit again corresponds to $\m \a 0, \alpha \a \infty$
keeping $\m e^{2\alpha}$ fixed. In this limit, $\l \a 1$ and
$h_{ij} \a \delta_{ij}$. So the extremal black hole metrics \newsoln\ and
\newfin\
are determined entirely by the function
$e^{2\phi}$. Since the solutions with and without a dilaton
differ only
by a factor of two in the exponent, {\it one can view the extremal
black hole with  dilaton as the square root of the extremal \RN\ solution.}
It is not clear what the physical interpretation of this is. On the
one hand,
the statement is clearly coordinate dependent. It holds in isotropic
coordinates where the spatial metric is manifestly conformally flat.
On the other hand,
the statement applies not just to single black holes but
extends to multi-black hole solutions as well. In the
Einstein-Maxwell theory, since there is no force between extremal black holes
(with the same sign of the charge), there exist static, multi-black hole
solutions. These are the Majumdar-Papepetrou solutions \mapa\ and take the form
\eqn\emmulti{ ds^2 = - e^{4\phi} dt^2 + e^{-4\phi} d \vec x \cdot d \vec x}
with
\eqn\multiphi{ e^{-2\phi} =1 + \Sigma_i
\frac{2M_i}{|\vec x - \vec x_i|}}
In exact analogy, the extremal black holes with a  massless
dilaton also have no force
between them. The fact that $Q^2 =2M^2$ rather than $Q^2 =  M^2$
in the extremal limit means that
there is a
stronger electrostatic repulsion which exactly cancels the additional
attractive
force due to the dilaton. The multi-black hole solution turns out
to be simply the ``square root"
of \emmulti:
\eqn\dilmulti{ ds^2 = - e^{2\phi} dt^2 + e^{-2\phi} d \vec x \cdot d \vec x}
with $e^{-2\phi}$ again given by  \multiphi.

The solution generating transformations \newem{}\ and \newemag{}\
produce solutions with electric
charge. But one can easily obtain solutions with magnetic charge by
applying a duality transformation. When the dilaton is present, the
duality transformation corresponds to replacing $F_{\mu\nu}$ and $\phi$
with  $\t F_{\mu\nu} \equiv \half e^{-2\phi}
\epsilon_{\mu\nu}^{~~~\lambda\rho}F_{\lambda\rho}$ and $\t \phi = -\phi$.
This leaves the stress energy tensor invariant and hence
does not change the Einstein metric. But since the dilaton changes
sign, it does change the string metric.  For electrically charged black
 holes, the string metric is obtained by multiplying \dilmulti\ by
$e^{2\phi}$. The result is that the spatial metric is completely
flat and $g_{tt}$ is identical to the Majumdar-Papepetrou solutions.
On the other hand, since $\phi$ changes
sign under duality, the string metric describing several magnetically
charged black holes is obtained by multiplying \dilmulti\ by $e^{-2\phi}$
where $\phi$ is still given by \multiphi. The resulting metric has
$g_{tt} = -1$ and a spatial metric which is identical to the
Majumdar-Papepetrou
solution.
In other words, the string metric describing several extremal black holes
is obtained from the analogous Einstein-Maxwell solution by flattening
either the time or space part of the metric (depending on the type of charge)
leaving the rest unchanged.

The magnetically charged case is of particular
interest.
Recall that in the spatial part of the \RN\ metric, the horizon moves off
to infinity as one approaches the extremal limit. The extremal geometry
resembles an infinite throat attached to an asymptotically flat region.
Since $g_{tt} = -1$ for the string solution, the horizon is now infinitely far
away in timelike and null directions as well as spacelike. It has been
suggested \banks\ that this type of geometry may play an important role
in explaining what happens to the information that falls into a black
hole, after it evaporates.

\newsec{Solutions with a Massive Dilaton}

We  now wish to include the effects of the dilaton mass.\foot{This section
is based work done in collaboration with J. Horne \hohomass. For another
discussion of black holes with a massive dilaton see \grha.} Unfortunately,
exact black hole solutions do not seem to be expressible in
closed form. Given the simple form of the \RN\ \rnyes\ and massless
dilaton  \einmet\ solutions, we will assume a metric of the form
\eqn\assume{ ds^2 = -\l dt^2 + \l^{-1}dr^2  + R^2 d\Omega}
where $\l$ and $R$ are functions of $r$ only.

We first consider the asymptotic form of the solution. We are interested
in solutions that are asymptotically flat, which requires that $\phi \a 0$
at infinity.
For large $r$, the right hand side of  the dilaton equation \eom{b}\
behaves like $Q^2/r^4$, and the
derivative term becomes negligible. The dilaton thus falls off like
$|\phi | \sim Q^2/m^2 r^4$. Recall that a massless scalar field falls off as
$1/r$ while a massive field with localized sources falls off exponentially.
Here we have a massive field with a source that falls off polynomially,
which results in the unusual asymptotic form of $\phi$. Note that
the limit $m\a 0$ keeping $r$ fixed is not well behaved. This is because
$|\phi |\sim Q^2/m^2 r^4$ only in
the asymptotic region where $r$ is large compared to the Compton
wavelength of the dilaton, i.e. $rm \gg 1$.

Now consider the metric equation. At large distances,
the Maxwell contribution to the stress tensor
will be $O(1/r^4)$ while all terms involving the dilaton will fall off much
faster. Thus, the asymptotic form of the field equation is identical to
the Einstein-Maxwell theory, and the solution is just \RN.
One can calculate the first order correction to the \RN\ solution by
treating the dilaton terms as a perturbation and one finds
\eqna\asym
$$\eqalignno{ \l & \sim 1- {2M\over r} +{Q^2 \over r^2} -
	 {Q^4 \over 5m^2 r^6} \>,  & \asym a\cr
	    R & \sim r \(1 -{2Q^4\over 7m^4 r^8}\) \>. & \asym b\cr }$$

Although the value of the dilaton mass is not known, one can place a lower
limit from the fact that the $1/r^2$ force law has been confirmed down to
scales of about $1$ cm. This requires that the dilaton Compton wavelength
be less than $1$ cm or $m> 10^{-5}$eV.  It follows that for a solar mass
black hole\foot{In geometrical units,
 the dilaton mass $m$ has dimensions of inverse length and is related to
 the inverse Compton wavelength. On the other hand, the black hole mass
 $M$ has dimensions of length and is related to the size of the black hole.
 Thus $Mm$ is dimensionless. Alternatively, one can view all quantities as
dimensionless and measured in Planck units.}, $Mm > 10^5$. In other words,
the black hole is much larger
 than the Compton wavelength of the dilaton.
 In this case, the deviation from \RN\ remains small until one is
 well inside the event horizon: At the horizon,
 $\delta \l < (Q/M)^4 (Mm)^{-2} \ll 1$
 and $\delta R < (Q/M)^4 (Mm)^{-4} \ll 1 $. So outside the horizon,
 the solution will be very similar to
 \RN. Since $\phi \sim Q^2/m^2r^4 < (Q/M)^2 (Mm)^{-2} \ll 1$,
 the dilaton remains small everywhere outside the event horizon.
 For $Q\approx M$, the inner horizon of \RN\ is close to the event horizon
 and the corrections due to the dilaton will still remain small there.
 So unlike the massless dilaton solution, a large
 black hole coupled to a massive dilaton can have two horizons. This is what
one
 should expect physically. When $Mm \gg 1$, the dilaton is essentially stuck
 in the bottom of its potential well and does not affect the solution
 significantly.

 Even though the dilaton does not qualitatively affect the
 geometry, it has an important consequence near the extremal limit.
 Like \RN, the horizons will coalesce
 in the extremal limit. However,
 the condition for when this occurs is no longer exactly $Q^2 = M^2$.
 The new condition is
 found by asking when $\l$ has a double zero. If we substitute
 the unperturbed condition, $Q^2=M^2$ and $r=M$, into the correction term
 in \asym{a}\ we
 find that to first order in the dilaton, $\l=0$ is equivalent to
 \eqn\extlmt{ r^2 - 2Mr + \(Q^2 -{1\over 5m^2}\) = 0 }
 This will have a double zero when
 \eqn\imp{M^2 = Q^2- {1\over 5m^2}}
 So $M^2 <Q^2$ in
 the extremal limit, just like the massless dilaton solutions. But at
 infinity, we have seen that the solution always approaches the \RN\
 solution, for which objects with $Q>M$ are repulsive. Since $Q^2$ is
 strictly  greater than $M^2$ in the extremal limit, nonextremal
 black holes can also have a charge greater than their mass. We conclude that
 {\it nearly extremal and extremal black holes with a massive dilaton
 are repulsive.} Roughly speaking, this is a result of the fact that
 the presence of the dilaton near the horizon allows $Q^2>M^2$ but the
 dilaton mass cuts off the attractive dilaton force at large separations.
This appears to be the first example of gravitationally bound repulsive
objects.

 Since the extremal black hole has a degenerate horizon, its Hawking
 temperature vanishes. However, this does not guarantee that it is stable.
 Extremal black holes might quantum mechanically bifurcate.
In the context of the Einstein-Maxwell theory, this was considered by
Brill \brill. He found an instanton describing the splitting of an extremal
\RN\ throat, and argued that there should exist a similar
finite action instanton
describing the splitting of the extremal \RN\ black hole. This has also
been considered in the theory with a massless dilaton \klopp. However
in both of these theories, the total mass $M$ of the black hole is
proportional to its charge $Q$ in the extremal limit. Thus one black hole
of charge $Q$ has the same mass as $n$ black holes of charge $Q/n$. The
solutions are degenerate. This is not the case for the theory with a
massive dilaton.
It is now energetically favorable
 for an extremal black hole to bifurcate. From \imp,
 the mass of a single extremal
 black hole of charge $Q$ is $M_1 = (Q^2 -
 1/5m^2)^{1\over 2}$ while the mass for $n$ widely separated extremal
 black holes with charge $Q/n$ is
 \eqn\separate{ M_n = n\[\({Q\over n}\)^2 - {1\over 5m^2} \]^{1\over 2}
 = \[Q^2 - {n^2\over 5m^2} \]^{1\over 2}  \>. }
 Clearly, $M_n$ is a decreasing function of $n$.
 These black holes should be even more likely to bifurcate than the ones
 considered previously.

The above expression for $M_n$ only applies to large black holes,
since it is based on \imp. (More precisely, it is valid when $mM_n/n \gg 1$.)
However the conclusion is likely to  hold more generally.
This is because it follows from the fact that extremal black
holes are repulsive, which in turn is a consequence of two general
properties of the extremal solutions:
they always approach \RN\ asymptotically and have
$Q^2 > M^2$. Thus extremal
black holes will probably continue to bifurcate until they have a single
unit of charge. At this point the black hole is small compared to the
Compton wavelength of the dilaton, $Mm \ll 1$. (Black holes of this type
can also be obtained by starting with a large hole with small charge, and
letting it evaporate.
The extremal limit will not be reached until $Mm \ll 1$.)
For these small black holes
there is a region $M \ll r \ll 1/m$ which is far from the black hole
but inside the dilaton Compton wavelength. One can solve the dilaton equation
\eom{b}\ exactly for $r \gg M$ since the spacetime is essentially flat.
The result is that for  $M \ll r \ll 1/m$, $\phi$ behaves like
a massless dilaton: $\phi \propto 1/r$. So one expects the black holes
to behave qualitatively
like the massless dilaton solution discussed in the previous section.
Numerical calculations confirm
that this is indeed the case \hohomass.
In particular, there is no inner horizon and in the extremal limit,
the event horizon shrinks to zero area and becomes singular.
The string metric describing an extremal magnetically charged black hole
has an infinite throat.

We have seen that black holes with a massive dilaton can be viewed as
interpolating between
the \RN\ solution (when their mass is large), and the massless dilaton solution
(when their mass is small).
One can say something about the transition region $Mm \approx 1$ by
considering when a degenerate horizon can exist. If there is
a radius $r_0$ for which $\l $ and $  \l'$ both vanish, the
field equations
yield the following condition on the value of $\phi$ at $r_0$
\eqn\dcon{  {e^{2\phi} \over \phi (1+ \phi)^2} =  Q^2 m^2 \>.}
The left hand side has a minimum of $e^2/4$ when $\phi = 1$. So for
$Q^2m^2 <e^2/4$, there can not exist a degenerate horizon. For
$Q^2m^2 >e^2/4$ there are two possible values of $\phi$, but the equations of
motion also imply that $\l'' \propto (1-\phi)$
so one needs $\phi <1$ to have $\l'' >0$ as required for an event horizon.
At the critical value, $Q^2m^2 =e^2/4$, $\l''$ also vanishes yielding
a triple horizon.
This suggests that
there may be solutions with three  distinct horizons. But at the moment,
there is no direct
evidence for this.

These intermediate size black holes can have an unusual property when
described in terms of the string metric. One can show that the radius of
the spheres of spherical symmetry in the Einstein metric, $R$,
is monotonically increasing outside the horizon as expected.
In the string metric, the radius of the spheres
is $Re^\phi$. For an electrically charged hole, it turns out that $\phi$
is also increasing outside the horizon so the spheres indeed
become larger as $r$
increases. However, for a magnetically charged hole, $\phi$ changes sign
and is monotonically decreasing outside the horizon.
So $Re^\phi$ need not be an increasing function of $r$. It certainly increases
near infinity, and for the massless dilaton solution \einmet,
it is monotonically
increasing everywhere. However, for nonzero dilaton mass, one can show that it
decreases outside the horizon for nearly extremal black holes with $Mm \approx
1$. This means that the spatial geometry contains a wormhole. Unlike
the familiar wormhole in the maximally extended Schwarzschild solution,
this wormhole is outside
the horizon. It is static and  transversable.

Combining this with our previous
results, we see that the string metric describing an extremal, magnetically
charged black hole has a rather exotic dependence on mass. It resembles
the \RN\ metric with  a degenerate
horizon when $Mm \gg 1$. A wormhole forms outside the degenerate horizon
when $Mm \approx 1$. Finally, the horizon and wormhole both disappear and
are replaced by an
infinite throat when $Mm\ll 1$.

\newsec{Conclusions}

We have discussed two related issues.
On the mathematical side, we have seen that despite the
apparent complication introduced by a massless dilaton, the theory
shares some of the features of the Einstein-Maxwell theory. In particular,
there is a solution generating transformation which is remarkably similar
to the Harrison transformation. The resulting
extremal black hole can be viewed as
the ``square root" of the extremal \RN\ solution.
On the physical side, we have seen
that the evolution of charged black holes predicted by string theory
is quite different from standard lore. Black holes initially (when $Mm \gg 1$)
are quite similar to the \RN\ solution. But as they evolve (through
Hawking evaporation and possible quantum bifurcation)
they reach a stage where $Mm \ll 1$.  At this point, the black holes are
similar to the massless dilaton solutions.
For a dilaton mass
of $m = 1$TeV, the transition will occur at a black hole mass of
$M=10^{11}$gms which
is well above the Planck scale. This ensures that Planck scale corrections,
which have been neglected throughout, will still be negligible.

What is the most likely endpoint of this evolution? Assuming the charge is
not completely radiated away, one expects to be left with an extremal black
hole with unit charge\foot{Although likely, this is still uncertain
due to the fact that the Hawking temperature does not vanish as one approaches
the extremal limit of these small black holes.}.
Since charge is quantized, this must be stable against
further bifurcation. These remnants would be pointlike objects, with charge
greater than their mass, which repel each other. In other words, they
would appear very much like  elementary particles. (See also \howi.)
In fact, by shifting
the dilaton potential so that its minimum is at $\phi_0 \ne 0 $,
one  can change the charge to mass ratio of the extremal black holes.
There even exists a value of $\phi_0$ such that the extremal
black hole of unit charge has the mass of an electron. In the past,
one of the main objections to interpreting an electron as a black hole
was the fact that its charge is much larger than its mass (in geometrical
units) so that according to the \RN\ solution, it would be a naked singularity.
The massive dilaton predicted by string theory removes this objection
and opens up the possibility of a much closer connection between
elementary particles and black holes.

\bigskip
\centerline{Acknowledgments}
It is a pleasure to thank my collaborators D. Garfinkle, A. Strominger,
and especially J. Horne.
This work was supported in part by NSF
Grant PHY-9008502.

\listrefs
\end